\documentclass[aps,twocolumn,showpacs,amsmath,amssymb,superscriptaddress,prc]{revtex4-2}
\usepackage{graphicx,here}
\usepackage{multirow}
\usepackage{tikz}
\usepackage{epstopdf}
\usepackage[percent]{overpic}
\usepackage{scrextend}
\usepackage{xcolor,xspace}
\usepackage[ulem=normalem]{changes}
\usepackage{hyperref}
\hypersetup{colorlinks=True,urlcolor=blue,linkcolor=blue,citecolor=blue,filecolor=black}

\colorlet{Changes@Color}{red}
\usepackage{dcolumn,color,footnote,bm,braket}
\usepackage{url,longtable,tabularx}
\usepackage{threeparttable}

\usepackage{fancybox}
\usetikzlibrary{matrix}

\newcommand\+{\dagger}
\newcommand\bqt{\tilde{\beta}_{2}}
\newcommand\bht{\tilde{\beta}_{4}}
\newcommand\bq{\beta_{2}}
\newcommand\bh{\beta_{4}}

\begin{document}

\title{Impacts of hexadecapole deformations on the collective energy spectra of axially deformed nuclei}

\author{L. Lotina}
\email{llotina.phy@pmf.hr}
\affiliation{Department of Physics, Faculty of Science, 
University of Zagreb, HR-10000 Zagreb, Croatia}

\author{K. Nomura}
\email{nomura@sci.hokudai.ac.jp}
\affiliation{Department of Physics, 
Hokkaido University, Sapporo 060-0810, Japan}
\affiliation{Nuclear Reaction Data Center, 
Hokkaido University, Sapporo 060-0810, Japan}

\date{\today}

\begin{abstract}
The hexadecapole deformation, as well as the 
quadrupole one, influences the 
low-lying states of finite nuclei. 
The hexadecapole correlations are often 
overshadowed by the large quadrupole effects, 
and hence have not been much investigated. 
Here we address the relevance of hexadecapole 
deformations in the calculations of 
low-energy collective states 
of heavy nuclei, by using the theoretical  
framework of the self-consistent mean-field method 
and the interacting-boson approximation. 
The interacting-boson Hamiltonian 
that explicitly includes the quadrupole and hexadecapole 
collective degrees of freedom is specified by a choice 
of the energy density functional and pairing 
interaction. 
In an illustrative application to axially deformed Gd 
isotopes, it is shown that the inclusion of the 
hexadecapole degree of freedom does not affect 
most of the low-spin and 
low-lying states qualitatively, 
but that has notable effects in that 
it significantly improves 
the description of high-spin states 
of the ground-state 
bands of nearly spherical vibrational nuclei 
and gives rise to 
$K^\pi=4^+$ bands exhibiting strong $E4$ transitions 
in strongly deformed nuclei. 
\end{abstract}

\maketitle

\section{Introduction}

Deformation of the nuclear surface and the 
corresponding collective excitations are a 
prominent aspect of the atomic nucleus \cite{BM}. 
Among those collective excitation modes 
that correspond to positive-parity states of nuclei, 
the dominant and most studied 
is of quadrupole type, 
while much less attention has been paid to  
the next leading order, 
hexadecapole deformation. 
This is mostly because  
the effects of the hexadecapole correlations 
in nuclear low-lying states are 
often overshadowed 
by large quadrupole correlation effects. 
The hexadecapole correlations, 
nevertheless, have been shown to be 
present in many 
rare-earth \cite{hendrie1968,erb1972,wollersheim1977,ronningen1977,BM}, 
and actinide \cite{bemis1973,zamfir1995} nuclei, 
as well as some light ones \cite{gupta2020}, 
and have recently been found 
in exotic isotopes with an unusual proton 
to neutron number ratio \cite{spieker2023}. 
Notable hexadecapole effects in 
nuclear collective structure include the 
appearance of the low-energy $K^\pi=4^+$ bands 
and enhanced electric hexadecapole 
($E4$) transitions. 
Furthermore, recent hydrodynamic 
simulation has indicated that 
the hexadecapole deformation plays a role in 
modeling heavy ion collisions studied 
at the Relativistic Heavy Ion 
Collider \cite{ryssens2023}. 
In addition, various nuclear deformation effects, 
including that of hexadecapole type, 
should have influences on the predictions 
of the neutrinoless double decay matrix elements 
of open shell nuclei \cite{engel2017}.

It is, therefore, interesting and timely to  
study impacts of hexadecapole 
deformations on nuclear structure 
in a quantitative and systematic way, 
using a model that allows for an accurate description 
of excitation spectra and electromagnetic 
transition properties 
of low-lying collective states. 
Among other nuclear structure models, 
the interacting boson model (IBM) \cite{IBM} 
has been successful for phenomenological 
descriptions of low-energy collective excitations 
in medium-heavy and heavy nuclei. 
The basic assumption of the IBM is that 
the nuclear low-lying states are described 
in terms of $s$ and $d$ bosons, which reflect 
\cite{OAI} the collective monopole, 
$S$ (with spin $0^+$), 
and quadrupole, $D$ (spin $2^+$), 
pairs of valence nucleons, respectively. 
The IBM should have certain 
microscopic foundations 
on the underlying nucleonic dynamics, 
and attempts have been made to derive the model 
Hamiltonian from more microscopic nuclear structure 
calculations 
\cite{OAI,mizusaki1997,nomura2008,nomura2010}. 
In particular, a mapping technique has been 
developed \cite{nomura2008} that links 
the IBM to the framework of the nuclear 
energy density functional (EDF). 
This procedure has been successfully 
applied to describe quadrupole 
\cite{nomura2008,nomura2010,nomura2011rot,nomura2012tri} 
and octupole \cite{nomura2013oct,nomura2014} 
collective states.

In addition to $s$ and $d$ bosons, 
spin $4^+$, or $g$, bosons have often been 
considered in the IBM \cite{IBM}. 
The importance of $g$ bosons in describing 
spectroscopic properties of deformed nuclei has been 
addressed from various perspectives 
\cite{bohr1980,otsuka1981,otsuka1982,bes1982,vanisacker1982,otsuka1985,otsuka-sugita1988,devi-kota1990,kuyucak1994,zerguine2008,vanisacker2010}. 
Some of these earlier studies 
also concern the validity of the $sd$-IBM from 
a microscopic point of view, that is, 
the question as to whether the $g$ boson 
degrees of freedom are indispensable or not for 
a precise description of axially deformed nuclei 
\cite{bohr1980,otsuka1981,otsuka1982,bes1982,otsuka1985}. 
Along with the strongly deformed regions, 
it should be also of interest to investigate the 
significance of the hexadecapole correlation effects 
in those nuclei in nearly spherical vibrational 
and transitional regions.

In this article, we implement the hexadecapole 
($g$ boson) degree freedom in the IBM by means of 
the aforementioned mapping procedure \cite{nomura2008}, 
and demonstrate that the 
hexadecapole effects are present in the low-energy 
collective states of heavy nuclei in the nearly 
spherical vibrational region as well as in the 
strongly deformed region. 
As an illustrative example we focus on 
the isotopic chain of 
axially deformed $^{148-160}$Gd nuclei, 
which exhibits a manifest 
first-order shape phase transition from 
spherical to (quadrupole) deformed shapes \cite{cejnar2010}, 
and for which hexadecapole collectivity 
has also been suggested to emerge empirically.

%-----------------------------------------------------------
%       PESs
%-----------------------------------------------------------
\begin{figure}
\begin{center}
\includegraphics[width=\linewidth]{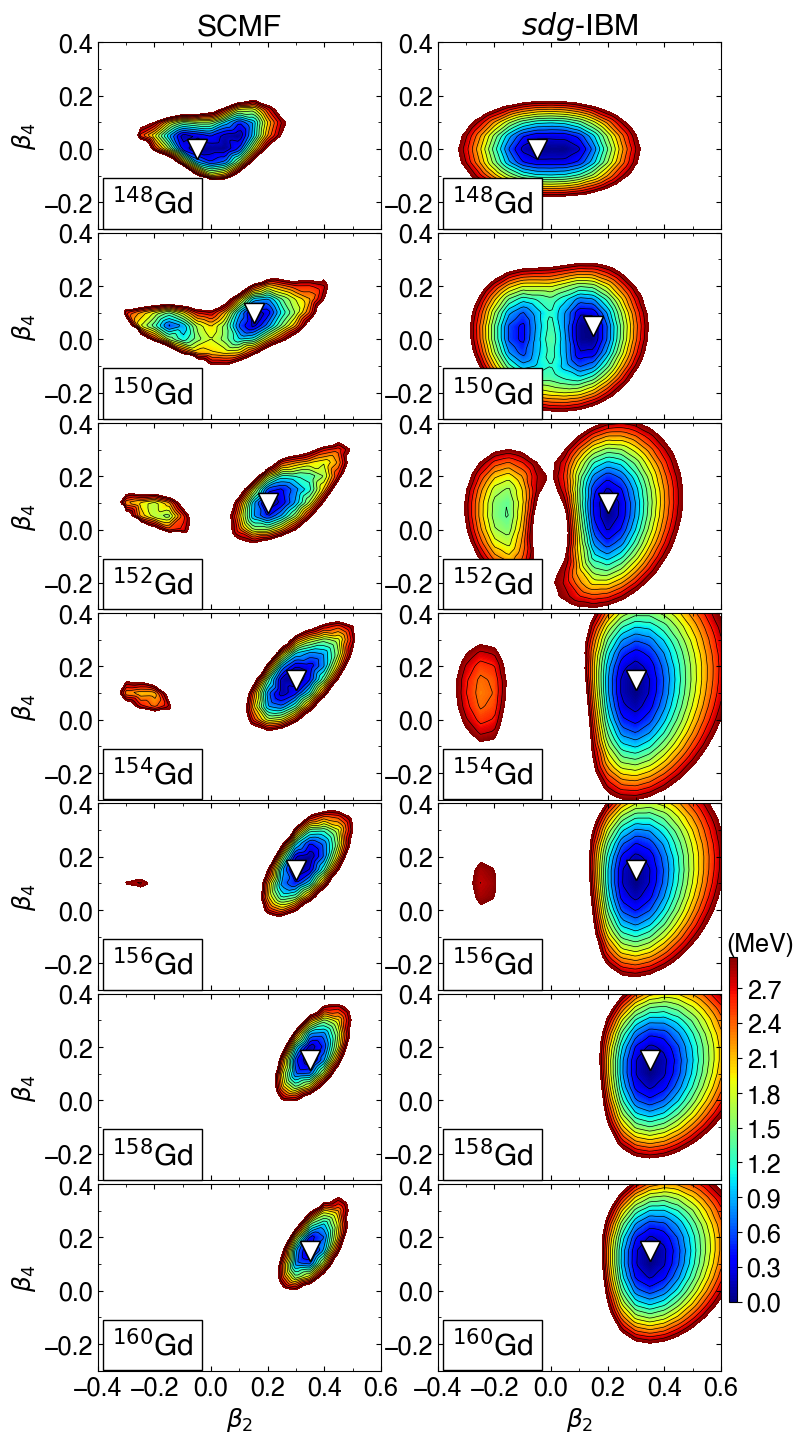}
\caption{Left column: axially symmetric quadrupole ($\bq$) and 
hexadecapole ($\bh$) 
constrained energy surfaces for the $^{148-160}$Gd 
isotopes calculated within the relativistic Hartree-Bogoliubov 
method using the density-dependent 
point-coupling energy density functional 
and the pairing force of finite range. 
Right column: 
the corresponding energy surfaces of the $sdg$-IBM. 
The global minimum is indicated by 
the open triangles.}
\label{fig:pes}
\end{center}
\end{figure}

\section{SCMF energy surfaces and mapping onto the IBM}

Our analysis begins with the 
self-consistent mean-field (SCMF) calculations 
using a nuclear EDF. 
The nuclear EDF approaches 
are nowadays among the most reliable theoretical 
methods of studying intrinsic and excited states 
of finite nuclei 
\cite{RS,bender2003,robledo2019,vretenar2005,meng2006,niksic2011,meng2015,zhou2016}, 
and hexadecapole deformations have also been 
considered as additional collective coordinates 
(see, e.g., Refs.~\cite{lalazissis1999,ryssens2023,kumar-robledo2023}). 
For the $^{148-160}$Gd isotopes, we perform 
the SCMF calculations within the 
multidimensionally constrained relativistic 
mean-field (MDC-RMF) model \cite{lu2014,zhou2016} 
and obtain energy surfaces 
in terms of the axial quadrupole ($\bq$) 
and hexadecapole ($\bh$) deformations, 
which are shown 
on the left column of Fig.~\ref{fig:pes}.  
The SCMF calculations are here carried out 
within the relativistic Hartree-Bogoliubov 
framework \cite{vretenar2005,niksic2011} 
using the density-dependent point-coupling (DD-PC1)
interaction \cite{DDPC1} and the separable 
pairing force of finite range \cite{tian2009}. 
The constraints are on the expectation 
values of axial quadrupole 
$\hat Q_{20}$ and hexadecapole 
$\hat Q_{40}$ moments, 
which are related to the deformation 
parameters $\beta_{2}$ 
and $\beta_{4}$ through the relation, 
$\beta_{\lambda}=({4\pi}/{3AR^\lambda})\braket{\hat Q_{\lambda0}}$
with $\lambda=2,4$ and $R=1.2A^{1/3}$ fm.

As one can see in Fig.~\ref{fig:pes}, 
for most of the Gd nuclei, 
the hexadecapole deformed ground state with 
a positive $\bh$ value is obtained in the SCMF 
energy surfaces: 
the global minimum occurs at the deformations 
(${\bq}^{\rm min}, {\bh}^{\rm min}$) 
$\approx$ ($-0.05$, 0), (0.15, 0.05), 
(0.2, 0.1), (0.3, 0.15), (0.3, 0.15), 
(0.35, 0.15), and (0.35, 0.15) for $^{148-160}$Gd, 
respectively. 
Both the ${\bq}^{\rm min}$ and ${\bh}^{\rm min}$ 
values keep increasing with the neutron number, 
but the latter changes more slowly 
than the former. 
One notices that there is no $\bh\neq0$ 
minimum on the energy surface of $^{148}$Gd. 
The potential is nevertheless  
rather soft along the $\bh$ direction, 
softest among the considered 
Gd isotopes. 
The softness in $\bh$ implies that 
the hexadecapole correlations 
play an important role in this nucleus, 
and, as we show below, to account for the 
$\bh$ softness the $g$ boson degree of freedom 
is required. 
These findings, regarding the 
$\bq-\bh$ energy surfaces, 
are consistent with earlier SCMF 
results obtained for the same mass region, 
e.g., the one with the axially deformed Woods-Saxon 
potential involving the hexadecapole degree 
of freedom \cite{nazarewicz1981}, and a 
more recent beyond-SCMF calculation that 
is based on the Gogny forces 
dealing with the quadrupole-hexadecapole 
coupling \cite{kumar-robledo2023}.

The SCMF results are then used to construct 
the Hamiltonian of the $s$, $d$, and $g$ 
boson system (denoted as $sdg$-IBM), 
which gives rise to excitation energies 
and electromagnetic transition rates.  
For the $sdg$-IBM Hamiltonian, we exploit 
the form that has been been shown to be adequate 
for phenomenological 
descriptions of shape phase transitions with 
quadrupole and hexadecapole degrees of freedom  
\cite{vanisacker2010}:
\begin{align}
\label{eq:ham-sdg}
 \hat H_{sdg} = \epsilon_d \hat n_d + \epsilon_g \hat n_g 
+ \kappa \hat Q \cdot \hat Q
+ \kappa (1-\chi^2) \hat Q' \cdot \hat Q' \; ,
\end{align}
where the first and second terms 
in the right-hand side stand for  
the $d$-, and $g$-boson number operators, 
$\hat n_d = d^\+ \cdot \tilde d$ 
and $\hat n_g = g^\+ \cdot \tilde g$, respectively. 
The third term represents 
the quadrupole-quadrupole interaction. 
The quadrupole operator, $\hat Q$, 
takes the form
 \begin{align}
  \hat Q
=&s^\+\tilde d + d^\+ s
+\chi
\biggl[
\frac{11\sqrt{10}}{28}(d^\+\times\tilde d)^{(2)}
\nonumber\\
&- \frac{9}{7} (d^\+\times\tilde g + g^\+\times\tilde d)^{(2)} 
+ \frac{3\sqrt{55}}{14} (g^\+\times\tilde g)^{(2)}
\biggr] \; ,
 \end{align}
which is the expression considered also in 
Ref.~\cite{vanisacker2010} and corresponds to 
a generator of $sdg$-SU(3) in the limit $\chi=1$ 
\cite{vanisacker2010,kota1987}. 
The last term on the right-hand side 
of Eq.~(\ref{eq:ham-sdg}) 
represents the hexadecapole-hexadecapole 
interaction, with the hexadecapole operator being 
$\hat Q' = s^\+\tilde g + g^\+s$. 
%
%------- added text ------------------------
In principle, the hexadecapole operator in 
the $sdg$-IBM could take a more complicated form 
that contains some other terms. 
The reason why we end up with the simplified form that 
comprises the $s^\+\tilde g + g^\+s$ terms only 
is because the SO(15) symmetry is assumed 
on the $sdg$-IBM Hamiltonian (see 
Refs.~\cite{devi-kota1990,vanisacker2010} for 
the details).  
In addition, for the sake of simplicity 
no distinction is made between the neutron 
and proton degrees of freedom. 
While a more realistic study would require 
that neutron and proton bosons be treated separately, 
it is expected that there is no qualitative 
difference in the description of the low-lying 
yrast states of most of the medium-heavy and heavy 
nuclei between the IBM that is comprised 
of neutron and proton bosons and the one in 
which they are not distinguished. 
The distinction between neutron and proton 
bosons would be relevant when describing 
phenomena such as the neutron-proton mixed 
symmetry states and the related magnetic 
dipole properties, in which the neutron 
and proton degrees of freedom play an 
important role. 
On the other hand, 
the scope of present work is to 
study the effect of $g$ bosons 
on energy spectra, and for 
the initial application of the mapped 
$sdg$-IBM framework to realistic cases, 
it would be sufficient 
to use a simpler version of the $sdg$-IBM, 
where the neutrons and protons are not 
distinguished. 
%-------------------------------------------

The parameters of the Hamiltonian (\ref{eq:ham-sdg}) 
($\epsilon_d$, $\epsilon_g$, 
$\kappa$ and $\chi$) are determined, 
for each nucleus, by applying the method 
of Ref.~\cite{nomura2008}: 
the $\bq-\bh$ SCMF 
energy surface, $E_\textnormal{SCMF}(\bq,\bh)$, 
is mapped onto the equivalent energy surface 
of the boson system, $E_\textnormal{IBM}(\bqt,\bht)$, 
so that the approximate equality, 
\begin{eqnarray}
 E_\textnormal{SCMF}(\bq,\bh) \approx E_\textnormal{IBM}(\bqt,\bht) \; ,
\end{eqnarray}
should be satisfied in the neighborhood of the 
global minimum. 
Here, $E_\textnormal{IBM}(\bqt,\bht)$ 
is given as the expectation value 
of the Hamiltonian (\ref{eq:ham-sdg})
in the coherent state $\ket{\phi}$, 
with $\ket{\phi} \propto
(1 + \bqt d^\+_0 + \bht g^\+_0)^{N_\textnormal{B}} \ket{0}$ 
\cite{ginocchio1980,devi-kota1990}. 
$N_\textnormal{B}$ stands for the number of bosons, 
which is equal to half the number of valence nucleons, 
and the ket $\ket{0}$ 
represents the inert core, i.e., 
the doubly magic nucleus $^{132}$Sn. 
The amplitudes $\bqt$ and $\bht$  
stand for the boson analogs of the 
axial quadrupole and 
hexadecapole deformations, respectively, 
and are assumed to be proportional to 
the fermionic counterparts, 
that is, $\bqt \propto \bq$ and 
$\bht \propto \bh$. 
See Refs.~\cite{nomura2008,nomura2010} 
for further details of this mapping procedure 
in the case of $sd$-IBM.

On the right column 
of Fig.~\ref{fig:pes}, we show the mapped $sdg$-IBM 
$\bq-\bh$ deformation energy surfaces. 
One can see that the basic characteristics  
of the SCMF energy surface, such as the depth 
of the potential and the coordinates corresponding 
to the global minimum 
[(${\bq}^{\rm min},{\bh}^{\rm min}$)], 
are reproduced in the 
mapped $sdg$-IBM surfaces.

To compare with the $sdg$-IBM results, 
we also carry out the calculations within 
the original version of the IBM, 
that comprises $s$ and $d$ 
bosons only ($sd$-IBM). 
The $sd$-IBM Hamiltonian here 
takes the standard form \cite{IBM}
\begin{eqnarray}
\label{eq:ham-sd}
 \hat H_{sd} = \epsilon_d \hat n_d + \kappa \hat Q_{sd} \cdot \hat Q_{sd} \; ,
\end{eqnarray}
where $\hat Q_{sd}=s^\+\tilde d + d^\+s 
+\chi_d(d^\+\times\tilde d)^{(2)}$ 
is the quadrupole operator for the 
($s,d$)-boson systems.  
The three parameters, 
$\epsilon_d$, $\kappa$, and $\chi_d$, 
are determined by mapping the 
SCMF energy surface along the $\bq$ deformation 
with $\bh=0$ onto that of the $sd$-IBM, that is, 
$E_\textnormal{SCMF}(\bq,\bh=0) \approx E_\textnormal{IBM}^{sd}(\bqt)$.

%-----------------------------------------------------------
%
%       Energies for g.s. band
%
%-----------------------------------------------------------
\begin{figure}
\begin{center}
\includegraphics[width=\linewidth]{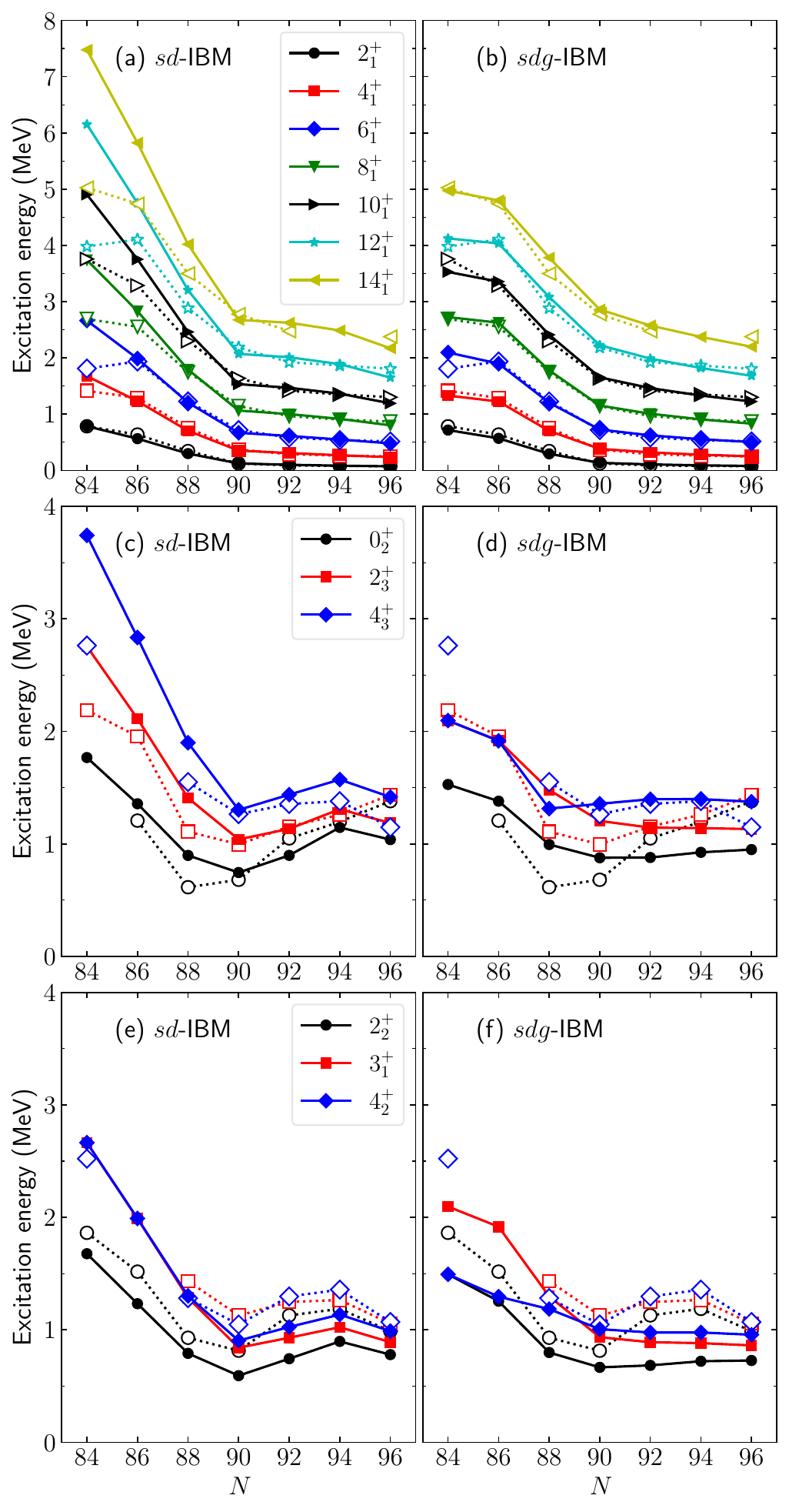}
\caption{Energy spectra of positive-parity 
even-spin yrast states [panels (a) and (b)], 
$0^+_2$, $2^+_3$, and $4^+_3$ states 
[panels (c) and (d)], 
and $2^+_2$, $3^+_1$, and $4^+_2$ states 
[panels (e) and (f)] 
of $^{148-160}$Gd calculated with the $sd$- 
(left column), and $sdg$-IBM (right column) in 
comparison with the experimental data \cite{data}. 
The calculated excitation energies are 
represented by filled symbols, and the 
corresponding experimental data by 
the open symbols.}
\label{fig:energies}
\end{center}
\end{figure}

\section{Results}

Figure~\ref{fig:energies}(a) and 
\ref{fig:energies}(b) show 
the excitation energies of 
even-spin positive-parity 
states in the ground-state bands of $^{148-160}$Gd, 
obtained from the diagonalization \cite{arbmodel} 
of the mapped $sd$- and $sdg$-IBM Hamiltonians, 
respectively, compared to the corresponding  
experimental data \cite{data}. 
A noticeable influence of including the hexadecapole 
deformation on the ground-state bands is that the excitation 
energies of the states with spin $I^\pi \geqslant 6^+$ 
calculated within the $sdg$-IBM 
for those nuclei with $N=84$ and 86, which are close 
to the neutron magic number $N=82$, 
are much lower than in the $sd$-IBM, 
and are in agreement with experiment.

The calculated ratios of 
the first $4^+$ to $2^+$ excitation energies,  
$R_{4/2}=E_x(4^+_1)/E_x(2^+_1)$, 
obtained from the $sdg$ ($sd$) IBM are, 
1.86 (2.13), 2.15 (2.18), 2.38 (2.36), 
2.84 (2.83), 2.99 (3.08), 3.10 (3.24), 3.16 (3.23) 
for $^{148-160}$Gd, respectively. 
The observed $R_{4/2}$ ratios, 
on the other hand, are equal to 
1.81, 2.02, 2.19, 3.02, 3.24, 
3.26, and 3.32 for $^{148-160}$Gd, 
respectively \cite{data}. 
By comparing the theoretical and 
experimental $R_{4/2}$ ratios, it turns 
out that another significant impact of 
$g$ bosons is that the $sdg$-IBM reproduces 
the experimental $R_{4/2}$ ratio for 
the $^{148}$Gd nucleus, $R_{4/2}=1.81<2$. 
The experimental value of $R_{4/2}<2$ 
could be reproduced only by the 
inclusion of $g$ bosons, 
since it lowers the $4^+_1$ level to be close 
in energy to the $2^+_1$ one. 
The fact that the observed $R_{4/2}$ ratio 
for $^{148}$Gd
is lower than two also reflects, to a good extent, 
the contribution from the single-particle 
excitations, which appear to be effectively 
accounted for by the inclusion of $g$ bosons.

%-------- added text -----------------------
Even though $g$ bosons, as well as $s$ 
and $d$ bosons, are considered as collective 
in nature, the fact that the inclusion of 
$g$ bosons in the IBM significantly improves 
description of the observed energy ratio 
of $R_{4/2}<2$, as well as the fact that the 
ground-state yrast levels with $I \geqslant 6$ 
is reproduced quite well, indicates that 
$g$ bosons are considered as necessary 
building blocks to describe the low-energy 
excitations in the nuclei 
with $N=84$ and 86, where single-particle 
degrees of freedom come to play a role. 
We also note that both the $sd$-IBM and 
$sdg$-IBM have often been 
applied to nearly spherical and moderately 
deformed nuclei with $N$ being 
near shell closure, and has been shown 
to be valid in a number of phenomenological 
applications and in the microscopic 
considerations. 
%-------- added text -----------------------

The lowering of the higher spin levels 
of the ground-state bands for the nearly spherical 
nuclei is explained by the increasing 
$g$ boson contribution to the 
wave functions as a function of spin. 
The contribution of $g$ bosons to a 
given state is inferred from the 
expectation value, $\braket{\hat n_g}$, 
computed by using the $sdg$-IBM wave function 
of that state. 
For $^{148}$Gd, in particular, 
the states with spin 
$I^\pi=4^+_1$, $6^+_1$, $8^+_1$, and $10^+_1$ 
are shown to contain one $g$ boson, i.e., 
$\braket{\hat n_g} \approx 1$, and 
$\braket{\hat n_g} \approx 1.5$ for the 
$12^+_1$, and $14^+_1$ states.

Figures~\ref{fig:energies}(c) and 
\ref{fig:energies}(d) depict 
the $0^+_2$, $2^+_3$, and $4^+_3$ energy 
levels, which are supposed to be part of the 
excited $K^\pi=0^+$ band in the well 
deformed isotopes with $N \geqslant 90$. 
The inclusion of $g$ bosons has an 
effect of lowering the $2^+_3$ and $4^+_3$ 
energy levels for those nuclei with $N\leqslant 88$. 
In particular, for the $4^+_3$ states 
of $^{148}$Gd, $^{150}$Gd, and $^{152}$Gd 
the expectation values are calculated as 
$\braket{\hat n_g} \approx 1$. 
On the other hand, the hexadecapole deformation 
makes only a minor effect 
on the $0^+_2$ energy levels in general, 
as the expectation 
values $\braket{\hat n_g} \approx 0$. 
Overall it appears that 
the $sdg$-IBM does not improve the description of 
these non-yrast states, and that even the $sd$-IBM, 
i.e., the calculation without $g$ bosons, 
is able to reproduce the observed 
behaviors of these states rather well.

Figures~\ref{fig:energies}(e) and 
\ref{fig:energies}(f) show the excitation 
spectra of the $2^+_2$, $3^+_1$, and $4^+_2$ 
states, which are attributed to members 
of the $\gamma$ vibrational band. 
The $g$ boson effect appears to be minor in 
the description of the bandhead $2^+_2$ level, 
but significantly lowers the energies of the 
$3^+_1$ and $4^+_2$ states for those nuclei 
with $N\leqslant 88$, which are weakly 
quadrupole and hexadecapole deformed. 
Particularly at $N=84$, the $4^+_2$ level 
obtained from the $sdg$-IBM is so low 
in energy as to be close to the $2^+_2$ 
level, in comparison to experiment.

The discrepancies between 
the calculated and experimental energy spectra 
for nonyrast states, as observed in 
Figs.~\ref{fig:energies}(c), \ref{fig:energies}(d), 
\ref{fig:energies}(e), and \ref{fig:energies}(f), 
are not surprising, given that, 
unlike the conventional IBM calculations, 
the Hamiltonian parameters are here not obtained 
from experiment, but from the mapping of the 
$\bq-\bh$ SCMF energy surface computed with 
the EDF that is not tailored for particular nuclei.

%-----------------------------------------------------------
%
%       154Gd spectra
%
%-----------------------------------------------------------
\begin{figure}
\begin{center}
\includegraphics[width=\linewidth]{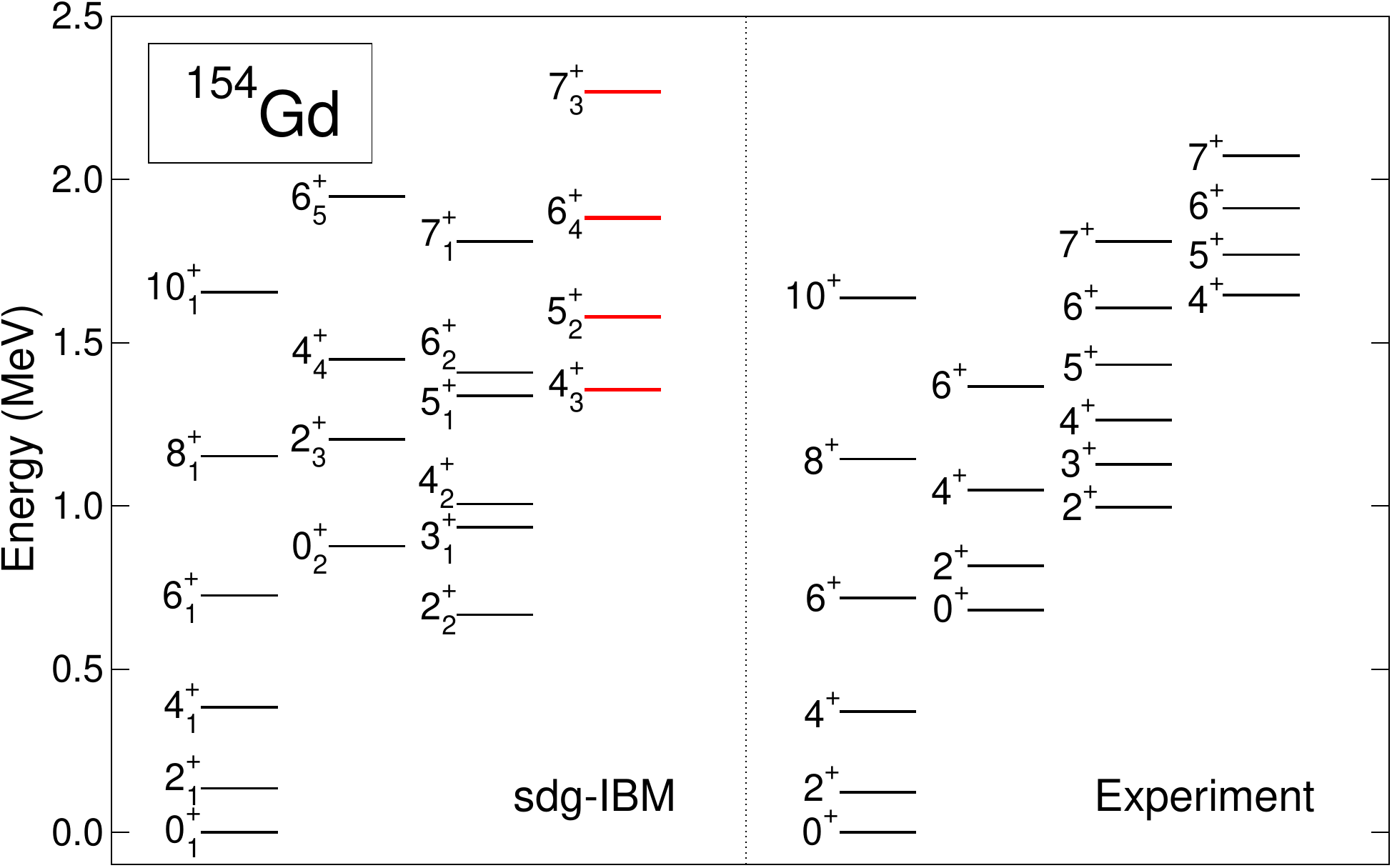}
\caption{Low-spin part of 
the positive-parity bands of $^{154}$Gd 
calculated with the $sdg$-IBM, in 
comparison with the experimental data \cite{data}. 
The theoretical band that is built on the 
$4^+_3$ state, highlighted 
in thick lines with color red, 
is predicted to be 
of one-$g$-boson character.}
\label{fig:gd154}
\end{center}
\end{figure}

The detailed band structure 
of each nucleus can also be studied. 
As a representative case 
we show in Fig.~\ref{fig:gd154} 
the low-energy positive-parity bands of 
$^{154}$Gd. 
In the $sdg$-IBM spectra, states are 
classified into bands according to 
the dominant inband $E2$ transitions and 
according to the nature of the states in terms of the 
$g$-boson content in their wave functions. 
One sees from Fig.~\ref{fig:gd154} 
that the ground-state band is reproduced well. 
The predicted $K^\pi=0^+_2$ band looks 
stretched in energy as compared to 
the experimental one, even though 
the $0^+_2$ bandhead energy is reasonably 
reproduced. 
The predicted $\gamma$ band, 
starting from the $2^+_2$ state, 
is much lower in energy 
than the experimental one. 
The appearance of the low-lying $2^+_2$ state 
indicates too pronounced $\gamma$ softness, 
which could be attributed to the 
particular choice of the nuclear EDF and 
pairing interaction.  
In addition, the calculated $\gamma$ band 
exhibits a staggering of levels, $(3^+_\gamma,4^+_\gamma)$, 
$(5^+_\gamma,6^+_\gamma)$, $\ldots$. 
It is a characteristic of a $\gamma$-unstable 
rotor \cite{gsoft}, but contradicts the 
observed feature of the $\gamma$ band, 
which looks rather harmonic. 
To remedy this, cubic terms are often included 
in the boson Hamiltonian 
\cite{nomura2012tri}, 
since they lower the energy levels of 
the odd-spin members of the $\gamma$ band, 
to be consistent with the observed 
$\gamma$-band structure that is 
harmonic.  
The cubic terms, however, would also 
lower the $2^+$ bandhead energy of 
the $\gamma$ band, which would become 
much lower than the experimental 
counterpart. 
Such an extension 
is well beyond the scope of this paper.

The theoretical band, built on top of the 
$4^+_3$ state, 
is interpreted as the $K^\pi=4^+$ band 
and is found to be of one-$g$-boson 
character in the present calculation. 
The experimental counterpart 
is the one with the bandhead energy 
$E_x(4^+_4)$ = 1646 keV \cite{data}. 
$E4$ transition properties 
are calculated 
with the transition operators that are given by
$e_{4,sdg}\left[\hat Q' + (d^\+ \times \tilde d)^{(4)}\right]$ 
for the $sdg$-IBM, and 
$e_{4,sd}(d^\+ \times \tilde d)^{(4)}$ 
for the $sd$-IBM. 
The addition of the $(d^\+ \times \tilde d)^{(4)}$ term 
to the $E4$ transition operator for 
the $sdg$-IBM is to compare the impact of $d$ 
bosons to that of $g$ bosons on the $E4$ 
transitions, and in that way one would be 
able to see properly how much the 
presence of $g$ bosons affects 
the $E4$ transitions. 
The effective boson charges, $e_{4,sdg}$ 
and $e_{4,sd}$, are 
fixed to reproduce an available experimental \cite{data}
$B(E4;4^+_1 \to 0^+_1)$ transition rate of 
38 $\pm$ 3 Weisskopf units (W.u.). 
The $sd$-IBM gives band structure 
of $^{154}$Gd qualitatively similar to 
that with the $sdg$-IBM. 
The $K^\pi=4^+$ band is also obtained 
in the $sd$-IBM with the bandhead $4^+_4$ 
state at the excitation energy of 1412 keV. 
A significant difference between the $sd$-
and $sdg$-IBM predictions is that the 
$B(E4;4_{K=4^+}^+ \to 0_{1}^+)$ transition 
obtained from the former is much 
lower (1.3 W.u.) than that from the latter (93 W.u.). 
Experiments to deduce the 
reduced $E4$ matrix elements, 
$|\braket{0^+_1\|M(E4)\|4^+_1}|$, were performed, 
e.g., in \cite{wollersheim1977,ronningen1977}, 
using the $(\alpha,\alpha')$ scattering and 
Coulomb excitations, and these transition 
matrix elements can be used to fix the 
boson effective charges, $e_{4,sd}$ and $e_{4,sdg}$. 
However, experimental information 
about other $E4$ transitions is not available 
for the Gd isotopes under study, and hence 
the detailed comparison of the present model 
with the experimental $E4$ transition rates 
is not feasible. 
An extensive study of the $E4$ properties 
in other isotopic chains will be reported 
elsewhere.

In the deformed region, the improvement of 
the $sdg$-IBM over the $sd$-IBM is also visible 
in the higher-lying states of a given spin. 
For even-spin states, for example, the sixth $2^+$ 
state and higher become lower and closer to 
each other in energy 
which is in a better agreement with experiment. 
For odd-spin states, this is even visible, 
e.g., for the third $3^+$ state and higher. 
However, the corresponding experimental data 
for the $3^+$ states are scarce. 
The $sdg$-IBM gives a slightly better description 
of the in-band $E2$ transitions, $I \to I-2$, 
within ground-state band for high-spin states, 
e.g., $I=8^+$ and $10^+$, for deformed Gd nuclei. 
There is, on the other hand, no qualitative difference 
in the calculated $E0$, as well as $E2$, transition 
properties for the low-spin states between the 
$sd$-IBM and $sdg$-IBM. 
These properties will be 
discussed in detail in a forthcoming longer article.

\section{Summary}

To summarize, we have analyzed the impacts of 
hexadecapole deformations on the low-lying 
collective states of axially symmetric 
heavy nuclei in the spherical vibrational 
as well as strongly deformed regimes. 
By using the results of the mean-field calculations 
based on the relativistic EDF, 
the $sdg$-IBM Hamiltonian has been determined 
without any adjustment to experiment. 
The inclusion of $g$ bosons has been shown to 
lower the states of ground-state bands with 
spin $I^\pi \geqslant 6^+$, especially in 
the region near the neutron closed shell $N=82$, 
thus improving the description of 
vibrational nuclei. 
For those nuclei with large quadrupole deformation, 
i.e., with $N \geqslant 90$, 
the $sdg$-IBM produces the $K^\pi=4^+$ band 
of one-$g$-boson character, 
which exhibits a much larger 
$B(E4;4^+_{K^\pi=4^+} \to 0^+_1)$ transition 
than in the $sd$-IBM. 
On the other hand, 
the $g$-boson effects on low-spin 
non-yrast states have 
been shown to be marginal 
for most of the deformed nuclei, in which cases 
the $sd$-IBM appears to reproduce 
the experimental data rather well. 
Now that we have a way of incorporating 
the quadrupole and hexadecapole degrees 
of freedom in the IBM in a unified manner, 
it can be applied to identify regions of the 
nuclear chart, 
including the experimentally unexplored ones, 
in which hexadecapole correlations 
may play prominent roles.

\acknowledgments
The work of L.L. is financed within 
the Tenure Track Pilot Programme of 
the Croatian Science Foundation and 
the \'Ecole Polytechnique F\'ed\'erale de Lausanne, 
and the project TTP-2018-07-3554 
Exotic Nuclear Structure and Dynamics, 
with funds from the Croatian-Swiss Research Programme. 

\bibliography{refs}

\end{document}